# High-resolution tomographic reconstruction of optical absorbance through scattering media using neural fields

## Author Information

Wuwei Ren[1,*,#], Siyuan Shen[1,#], Linlin Li[1,#], Shengyu Gao[1], Yuehan Wang[1], Liangtao Gu[1], Shiying Li[1], Xingjun Zhu[2], Jiahua Jiang[3,4,*], Jingyi Yu[1,*]

[1]*School of Information Science and Technology, ShanghaiTech University, Shanghai 201210, China*

[2]*School of Physical Science and Technology, ShanghaiTech University, Shanghai 201210, China*

[3]*Institute of Mathematical Science, ShanghaiTech University, Shanghai 201210, China*

[4]*School of Mathematics, University of Birmingham, Edgbaston, B15 2QN, United Kingdom*

[#]*Equal contribution*

[*]*Corresponding author*

## Corresponding authors

Correspondence to:

Jingyi Yu (yujingyi@shanghaitech.edu.cn), Wuwei Ren (renww@shanghaitech.edu.cn)

School of Information Science and Technology, ShanghaiTech University, Shanghai 201210, China

Jiahua Jiang (j.jiang.3@bham.ac.uk)

School of Mathematics, University of Birmingham, Edgbaston, B15 2QN, United Kingdom

## Abstract

Light scattering imposes a major obstacle for imaging objects seated deeply in turbid media, such as biological tissues and foggy air. Diffuse optical tomography (DOT) tackles scattering by volumetrically recovering the optical absorbance and has shown significance in medical imaging, remote sensing and autonomous driving. A conventional DOT reconstruction paradigm necessitates discretizing the object volume into voxels at a pre-determined resolution for modelling diffuse light propagation and the resulting spatial resolution of the reconstruction is generally limited. We propose NeuDOT, a novel DOT scheme based on neural fields (NF) to continuously encode the optical absorbance within the volume and subsequently bridge the gap between model accuracy and high resolution. Comprehensive experiments demonstrate that NeuDOT achieves submillimetre lateral resolution and resolves complex 3D objects at 14 mm-depth, outperforming the state-of-the-art methods. NeuDOT is a non-invasive, high-resolution and computationally efficient tomographic method, and unlocks further applications of NF involving light scattering.

# Introduction

The capability of non-invasively probing optical properties including absorption and scattering coefficients through highly diffusive media, such as biological tissues, foggy air, and murky water, can potentially benefit numerous engineering and research disciplines. A classic example is the routinely used pulse oximetry during the COVID pandemic, which emits near-infrared light to monitor the oxygenation level based on a simple mechanism: the absorption coefficients at multiple wavelengths differs significantly between blood saturated with oxygen and that lacking oxygen. The one-dimensional sensing technique of oximetry has since been further extended to more challenging three-dimensional imaging tasks, ranging from volumetric rendering of vascular anatomy in highly scattering tissues to accurate localization of pedestrians through fog for autonomous driving[1,2]. For many biomedical applications in particular, the tissue viability and functionality can be more clearly manifested by using 3D hemodynamic signatures than a one-dimensional sensing technique like oximetry.

Diffuse optical tomography (DOT) has emerged as an effective three-dimensional imaging technique with the capability of volumetrically mapping the optical properties in the depth up to a few centimetres within a scattering medium[3]. In a nutshell, DOT non-invasively measures the transmitted and scattered light from the object surface, and subsequently applies a model-based reconstruction algorithm to recover a 3D map on the optical properties, in particular optical absorption ($\mu_a$), within the object volume. DOT has demonstrated significant values in breast cancer diagnostics[4,5], functional brain imaging[6,7], skeletal muscle assessment[8], and peripheral circulation evaluation[9]. More recent advances attempt to fine-tune DOT to image objects of a much larger scale (e.g., meter-sized), with potential applications in remote sensing and autonomous driving[10].

Despite its abundant potential applications, the image reconstruction of DOT constitutes a fundamental challenge. Conventional DOT reconstruction methods lead to low resolution ranging from 1 mm to 1 cm[11], restricting its wider deployment in the scenarios that requires detection of subtle abnormalities in the tissue. The causes are multi-fold. Firstly, DOT combats with multiply-scattered photons and renders its inverse problem highly ill-posed[12,13]. Light scattering prevents an imaging system from accurately focusing on the target location and also incurs severe blurring in the imaging plane, especially when the object depth exceeds the detection range of canonical microscopic techniques, i.e., at maximally ~1 mm, also known as the transport mean free path (TMFP)[1,14]. Therefore, the measurement system is non-linear and classic reconstruction algorithms for X-ray computed tomography (CT), such as filtered back-projection (FBP), is not applicable in DOT, as those algorithms merely accounts for ballistic or low-scattered photon travelling[15]. Secondly, high-resolution DOT reconstruction is computationally expensive. A high-precision reconstruction algorithm requires accurate modelling of diffuse light propagation, which can be numerically implemented by initially discretizing the object volume into voxels or mesh elements and then applying time-consuming finite element methods (FEM) or Monte Carlo (MC) methods[16,17]. However, for large-sized and optically heterogeneous objects, large and dense voxel or mesh approximation is demanded. Yet, discretization by adding more voxels or elements not only overburdens computational cost, but also incurs a plethora of the unknowns that aggravates the ill-posedness of the inverse problem[3]. Overall, developing a high-resolution DOT reconstruction algorithm that is capable of addressing the ill-posed nature of the inverse problem and reducing the computational cost is highly attractive.

To tackle these problems, a conventional paradigm of DOT reconstruction focuses on seeking a optimal solution of unknown optical properties (mainly $\mu_a$) by minimizing a penalty function consisting of a data-fidelity term and a regularization term. The former represents the discrepancy between the real measurement and the prediction via a physics-based diffusion model, whilst the latter introduces prior knowledge regarding the solution, such as restriction of smoothness or sparsity[8]. A combination of a least-square loss and a Tikhonov regularization with a $l_2$ norm can significantly reduce the illposedness of the DOT inverse problem. Nevertheless, undesirable over-smooth effects is presented in the reconstructed image[7]. Apart from the conventional reconstruction, deep learning (DL) approaches have demonstrated better accuracy in DOT reconstruction[18]. The recent renaissance of these DL approaches has been mainly driven by convolutional neural network (CNN)-based methods. For example, a typical CNN architecture first learns the non-linear physics of the inverse scattering problem and then recovers heterogenous optical maps in biomimetic phantoms and live animals[19]. A multitask learning network has been deployed to reconstruct the location of lesions in a limited-angle DOT setting[20]. However, nearly all existing DL approaches are based on a supervised strategy: to obtain high-resolution reconstructions, these end-to-end DL-based methods require a large amount of training datasets produced either by real experiments or expensive simulations. Unsupervised or self-supervised DL approaches are hence in great need, without the necessity of extensive training dataset generation. It is also worth noting that both conventional and DL-based DOT reconstruction methods by far require inevitable steps of discretizing the object volume first and subsequently seeking a discretized version of the solution. Nevertheless, as mentioned before, refinement of the discretized voxels or mesh elements are linked with the rise of computational cost and ill-posedness. Hence, there is an urgent need to shift from discretization-based representation to continuous presentations suitable for self-supervised DL for supporting high-resolution DOT imaging.

Neural Radiance Field (NeRF) emerges as a new self-supervised DL scheme that can circumvent the discretization-based representation and continuously renders a volumetric target via a fully-connected deep network[21]. Inspired by the original application of NeRF in view synthesis, neural field (NF)-based reconstruction methods has been successfully used in several imaging techniques, including non-line-sight imaging (NLOS)[22], microscopy[23,24], and limited-projection computed tomography (CT)[25], demonstrating that NF is powerful in learning a high-quality representation of the targeted object from insufficient measurement (e.g., 2D projections from limited angles in CT) without external training datasets. However, most of these NF-based reconstruction methods only involve the ballistic photon propagation, which can be treated as a straightforward extension of NeRF in its original application of view synthesis[21]. NF-based rendering or reconstruction involving severe light scattering is less reported. Herein we propose neural field-based diffuse optical tomography (NeuDOT), a novel DOT reconstruction scheme for learning a continuous, high-fidelity and 3D absorption coefficient distribution, or $\mu_a$ map, from a series of 2D measurements without an external training procedure. Our method features several unique advantages:

- Our approach inherits the prominent benefits of NF-based representation: its continuous representation allows us to have sufficient resolution to pinpoint important areas at arbitrary positions. This is achieved by optimizing a multi-layer perceptron (MLP) neural network with the aim to learn an implicit function mapping the spatial coordinates to the $\mu_a$ map. NeuDOT can efficiently store and visualize a high-resolution reconstruction result, outperforming the conventional methods with an explicit, pre-determined mesh or voxel grid.

- NeuDOT is the first exploration of a NF-based rendering technique that enables imaging

through highly scattering media. An FEM simulator is seamlessly integrated into the NF-based reconstruction scheme, which can accurately predict diffusive light propagation and accounts for optical heterogeneity by assigning different absorption and scattering coefficients to each element of the FEM mesh.

- NeuDOT is computationally efficient in aspects of reconstruction speed and memory usage. Our method is self-supervised, thus no time-consuming external training is needed. To further alleviate the computational cost, we introduce adaptive meshing by refining the regions of interests in the FEM mesh during iterations of reconstruction. New nodal values of $\mu_a$ in an updated mesh can be retrieved by querying the trained MLP. Therefore, the accuracy of the forward model increases, but the unknown problem size (i.e., the structure of the MLP) remains the same.

Comprehensive experiments demonstrate that our NeuDOT outperforms widely used conventional methods as well as the latest end-to-end DL methods by achieving unprecedented spatial resolution in phantom studies with different levels of structural complexities.

## Results

### System description and NeuDOT reconstruction

We have devised a continuous wave (CW)-DOT imaging system (Fig. 1a) equipped with a solid-state laser and a 16-bit CMOS camera providing a field-of-view (FOV) of $60 \times 60$ mm$^2$ on the customized sample stage. The generated free-space laser beam illuminates the bottom surface of the scattering sample with a raster scanning pattern steered by a galvanomirror, whilst the camera captures one image on the top surface at each scan $p_i$ ($i = 1,2,...N$, where $N$ denotes the total number of illumination points). The resulting 2D image stack (Fig. 1c) is used as the input for DOT reconstruction (see details in Methods). The target of DOT is to estimate the heterogenous optical coefficient $\mu_a$ from the boundary measurement $\Gamma$ (Fig. 1b). A widely used strategy to combat the ill-posedness of the inversion procedure is via optimization[3], e.g., by minimizing the following objective function,

$$\Psi(\mu_a) = \|\Gamma - f(\mu_a)\|_2^2 + \tau R(\mu_a) \qquad (1)$$

with forward model $f$ given by Eqs. (3)-(5), regularization function $R$ and hyperparameter $\tau$ (See detailed derivation of forward model in Methods).

The main bottleneck of reconstructing complex targets using classical DOT reconstruction is that the basic implementation for optimizing the mesh-based representation of $\mu_a(r)$ cannot easily reach sufficiently high resolution: increasing mesh-fineness of the forward model reduces the robustness and degrades the accuracy of the optimization solvers[26]. To improve the spatial resolution without degrading the robustness, we exploit aspects of both NF rendering and physics-based finite element model via a new computational framework for super-resolution DOT reconstruction that we call NeuDOT (Fig. 1b). NeuDOT aims to approximate the distribution of optical absorption with NF defined using an MLP network $\mu_a^\Theta(r)$ whose weights $\Theta$ are optimized by minimizing a modified objective function of Eq. (1), given by

$$\Psi(\mu_a^\Theta(r)) = \left\|\Gamma - f(\mu_a^\Theta(r))\right\|_2^2 + \tau R(\mu_a^\Theta(r)) \qquad (2)$$

Two key features of NeuDOT are crucial for high quality reconstruction: 1) The size of absorption distribution $\mu_a^\Theta(r)$ is independent of the finite element mesh, which enhances the robustness of our algorithm and heightens the possibility of attaining super-resolution reconstruction. 2) NeuDOT is unsupervised and therefore does not require traditionally large training datasets. The pipeline of NeuDOT contains two major steps: MLP construction and adaptive meshing (Fig. 1b). The comparison between Conventional DOT and NeuDOT is illustrated by Fig. 1d. (See details about MLP and adaptive meshing in Methods, Supplementary Movie 1).

Reconstruction performance characterization

We evaluated the reconstruction performance of NeuDOT starting from the lateral direction based on the measurement obtained from a scattering phantom containing different two-dimensional planar patterns (Experiment 1, see details in Methods). On all the acquired datasets, we compared NeuDOT over two state-of-the-art DOT reconstruction methods: 1) a classic reconstruction method using a physics-based model with modified Levenberg-Marquardt algorithm (LM)[27] (Supplementary Note 1) and 2) a recently proposed supervised learning method using a CNN[20] (Supplementary Note 2). Regardless of different imaging targets and reconstruction algorithms used, the results are stored in a grid of 110 × 110 × 7 voxels uniformly. To illustrate the difference from the ground truth, we display only the bottom layer (z = 0 mm) of the reconstruction results (Fig. 2). All three methods managed to reconstruct the shape of letter 'T' (Fig. 2a, row 1). For comparison, NeuDOT recovered sharper boundaries as well as more homogeneous intensity distributions over LM and CNN. The CNN results, in contrast, appear over-smoothed on the boundaries and at the same time exhibit strong noise in the background. In the case of university logo (Fig. 2a, row 2), NeuDOT separates the two surrounding rings as well as partially recovers the central tower pattern. Nevertheless, neither LM or CNN resolved the tower or the two rings. In the last case, NeuDOT clearly reconstructed the two stripes separated 0.5 mm apart while LM or CNN cannot (Fig. 2a, row 3). In addition, we also draw a horizontal profile cross the center in reconstruction results and the ground truth (Fig. 2c), showing that NeuDOT achieves sub-millimeter lateral resolution, much higher than LM or CNN. Quantitative comparisons of different metrics for different planar cases are shown in Table. S5 in Supplementary Note 7, demonstrating the advantages of NeuDOT over the SOTA.

Next, we assessed reconstruction quality along the axial direction by varying only the height of imaging targets in a scattering phantom (Experiment 2, see details in Methods). For all reconstruction methods, we compute the recovered absorption coefficients within a uniformly discretized volume of $110 \times 110 \times 15$ voxels. NeuDOT is capable of recovering the height of all three rods with higher accuracy (Figs. 3a and b). In comparison, LM fails to recover the depth information of the rods. In addition, the squre cross sections of each rod is well preserved with a clear boundary in NeuDOT, whereas those boundaries are distorted in LM and CNN. We have also compared profile drawings along x- and y-axes of each cross section (Fig. 3c). These profiles provide a clear visual representation of NeuDOT's ability to produce highly precise results relative to the SOTA techniques.

## Artificial blood vessel phantom

In the last experiment (Experiment 3, see details in Methods), we imaged a phantom that contains a virtual blood vessel structure extracted from a human cerebrovascular model, a relatively challenging but more realistic task for DOT imaging. Different cross sections of the reconstruction results (z = 3, 5, 9 mm) are displayed in Fig. 4c. LM can only resolve the first layer of the object, but fails to generate images with sufficient intensity in the remaining slices. CNN-based method can slightly improve the reconstruction quality, but produces excessively blurred images. In comparison, NeuDOT not only revolves the basic structures at all slices, but also recovers sub-branches of the vessel model (Fig. 4c). The capability of recovering multiple slices at different depths further enables NeuDOT to recover the 3D shape of the vessel, unprecedented in prior art. Specifically, we perform skeleton extraction using medial axis transform[28] and compare NeuDOT estimation with the ground truth (Fig. 4d). The depth of the extracted central line is colour encoded, whereas the joint and end points of the vessel branches are highlighted in red. NeuDOT is able to recover the basic anatomy of the vessel including major branches and joints. The length of minor branches on the NeuDOT results still differs from the original phantom, largely attributed to noise and insufficient observations. Further comparison of metrics are consistent with the visual reconstructions in all slices (Supplementary Note 7).

## Computational storage and time

Lastly, we evaluated the computational cost of NeuDOT by analyzing the memory cost and reconstruction time in comparison with other methods. For simplicity, the proof of concept is obtained from experiment 2 with three rods. We observe that NeuDOT requires least memory space (Fig. 3d), largely benefited from the step of adaptive meshing. The stair-step-shaped growth of storage usage rises from 4530 M to 19060 M at the iteration of 1200 times, as the FEM mesh is periodically updated by adding new sampling point with an interval of 100 iterations (Supplementary Note 3). We also tested the same NF-based method without adaptive meshing. To achieve equivalent reconstruction resolution, a fixed-dimension fine mesh is used and consequently gave rise to a significant growth of memory usage up to approximately 40000 M, doubling the usage in the case when using adaptive meshing. Another DL-based method, namely the end-to-end CNN method, is computationally efficient with memory cost of approximately 20000 M, close to the level of NeuDOT during the finishing stage of iterations. LM is most memory-consuming due to the requisite of saving enormous Jacobian matrix for dense FEM mesh.

Notice that NeuDOT without adaptive mesh is still more memory-economical than LM. Compared with NeuDOT and the CNN-based method, it still suffers onerous storage burden, which is the motivation for us to propose adaptive meshing. Here we have demonstrated that NeuDOT empowered by adaptive meshing capacitates the flexibility of its applications for diversified objects, different accuracy requirements and more general environment. For example, in the computational environment with limited storage, NeuDOT is still able to provide a relatively high-resolution reconstruction.

Regarding the aspect of efficiency, the computational time of each method for experiment 2 is plotted (Fig. 3e). Availed by the adaptive meshing, we see a factor of 3.65 times runtime speedup between NeuDOT (~60 min) and NeuDOT without adaptive mesh (~219 min). LM (~6600 min) is most expensive among these four approaches. In all of these cases, 1200 iterations are counted. Nevertheless, in practice, the second order convergence of LM allows us to get a satisfied reconstruction in only 10 iterations, which takes comparable time as NeuDOT (~60 min). For the CNN-based method, once the network training (210 min) is completed, the time for the reconstruction is in the order of a few tenths of seconds (~0.5 seconds). However, the training time highly depends on the amount of data, and insufficient data will stymie the generalization performance of CNN (Supplementary Note 2).

## Discussion

Resolving objects from severely scattering medium is a long standing challenge for various research areas, specially for biomedical research and autonomous driving. During the last decade, techniques such as multiphoton microscopy[29], photoacoustic tomography[30], and wavefront shaping[31] have improved the penetration depth of optical imaging by either manipulating the illumination or acquiring non-scattered transferred energy signal (e.g., using photoacoustic signal). Despite these attempts of technical improvement, the reconstruction paradigm of diffuse optical imaging remained unchanged: solving an inverse problem with a fixed-dimension unknown parameter, which largely prevents the improvement of spatial resolution. Developing a new reconstruction paradigm that empowers better imaging performance of an existing system without increasing the hardware complexity is in high demand. In this work, we shifted the classic DOT paradigm to NeuDOT, a new scheme in which the unknown optical property map is implicitly represented using a positional encoded NF containing high-frequency spatial information[21]. This reconstruction method was further validated experimentally using a homebuilt DOT system. We conducted a series of elaborated phantom experiments by increasing the complexity from a simple 2D object case to a more sophisticated 3D artificial vessel model. NeuDOT achieves submillimeter resolution for an embedded a planar object within a tissue-mimicking phantom of 6.5 mm thickness (Fig. 2b), and volumetrically resolves the complex structure of a virtual vessel model (Fig. 4). For all the experiments, NeuDOT outperforms both a classic LM reconstruction[27] and the state-of-the-art CNN method[19].

The novelty of our proposed method lies in the hybridization of the NF-based continuous representation of optical absorbance and the adapted physic-based light propagation modeling using FEM. Regarding the former part of NF, the first step of positional encoding maps the each position in 3D coordinates $(x, y, z)$ into a single variable of optical absorption coefficient $\mu_a^\Theta(r)$ using a series of trigonometric functions with increasing frequencies in Eq. (6). Such a step preserves not only the values of $\mu_a^\Theta(r)$ at any arbitrary position but also its high order derivative property due to the intrinsic property of trigonometric functions, i.e., the derivatives of a paired sine-cosine function remains unchanged. Next, the unknown $\mu_a^\Theta(r)$ is continuously represented using an MLP network with a large number of weighting parameters $\Theta$. Notably, the classic method like LM and the recently reported end-to-end CNN-based method only assumes a fixed-dimension solution[19,27]. NeuDOT achieves unprecedented spatial resolution in the reconstructed images, and largely removes the over smoothed effect resulted from the classic methods. The second part of physic-based forward modelling is also crucial. For each iteration, the simulated measurement generated from the FEM solver[32] which accounts for the optical heterogeneity is compared to the real experimental data. The discrepancy information is then used to update the MLP network, subsequently changing the values of $\mu_a^\Theta(\boldsymbol{r})$ to plug into the FEM solver in the next iteration. As a result, our method is unsupervised and does not require extra training step like CNN which is highly time consuming. A major obstacle limiting the model accuracy is that a dense mesh with a large number of nodes will cause long computational time and excessive memory cost. To make our algorithm more efficient, we introduced adaptive meshing by refining the regions with high optical heterogeneity (high absorption in our case). The implicit function is advantageous in that we can easily re-sample from it by arbitrarily giving any 3D coordinates to the MLP. Gibbs sampling is used to sample more rationally based on the intermediate reconstruction result, yielding new nodes added back to the existing mesh. The mesh is thus updated for the next iteration, refining the accuracy of forward modelling. It shows that our adaptive meshing strategy can improve the computational speed of more than 3 times and save the memory usage of about 2 times when compared with the same method without this crucial adaptive meshing step. Besides, the classic LM method is a second-order convergence algorithm, which requires fewer iterations and has demonstrated its robustness in solving various DOT reconstruction problems. Nevertheless, the disadvantage of LM is that it requires a large amount of memory for storing and operating the Jacobian matrix, which is unacceptable for neural networks with a large number of parameters.

Same as other NF approaches, NeuDOT uses a robust first-order Adam optimizer[33] since it is computationally efficient and proved working well on a series of NF applications[21].

The significance of NeuDOT also lies in that the proposed pipeline can be generalized to other diffuse optical imaging methods and even other imaging modalities that involves scattering effects. In this work, we used a CW light source and a CMOS camera for recording. This CW-DOT technique features high signal-to-noise (SNR) data quality, simplicity in the system configurations, and low cost, thus it is widely adopted for clinical and biomedical research, e.g., functional brain imaging[34]. A potential improvement of DOT technique is to introduce time-domain (TD) information instead of only photon intensity averaged in a certain exposure time as the current CW-DOT system[11]. Single photon avalanche diodes (SPAD) and picosecond pulsed lasers are combined to facilitate TD measurement generating much a large number of data, which has been demonstrated to significantly improve the DOT reconstruction quality[11,35]. Besides, instead of optical properties, extrinsic contrast agent such as fluorescent probes can be also volumetrically reconstructed using a similar principle of DOT[36,37]. For achieving TD-DOT and fluorescence reconstruction, NeuDOT can be easily adapted by changing the current FEM solver with additional variables, whilst the rest of the pipeline can be remained unchanged. In addition, we may translate the NeuDOT concept to other imaging modalities, e.g., positron emission tomography (PET)[38] and X-ray CT[39], as scattering photons should be considered to improve the reconstruction quality. NeuDOT can be also considered as the technical extension of original NF-based view synthesis which only accounts for ballistic photon propagation[21,40].

Despite the promising reconstruction results our approach achieves in the real experiments, this is merely the first step toward high-precision and high-speed 3D imaging in a scattering medium. The forward modeling module in NeuDOT is implemented with an FEM solver, but the FEM-based diffusion approximation collapses in the region close to the light source or in the case when $\mu_a \ll \mu_s'$ is not satisfied[16,41]. An alternative to improve simulation accuracy is by using a more precise and more time-consuming MC-based method[17], a forward model also compatible with our current NeuDOT pipeline. The current computational cost of NeuDOT still cannot fulfill the requirement of real-time 3D visualization of optical absorbance. Fortunately, subsequent work in the field of computer vision is reported to improve the disadvantage of high training cost of NF.[42,43,44,45] For example, instead of using purely MLP, Chen et al propose a 4D tensor to model the radiance field and further factorize the tensor into low-rank components to speed up the training process[42]. Müller et al attempt to reduce the computational cost by using a new input encoding strategy allowing a smaller network without sacrificing the image quality[43]. Except the phantom targets used in this work, we plan to use our NeuDOT method for more realistic and complicated cases, e.g., in vivo imaging involving animals or even human tissues. We firmly believe that our method can find wide applications at different scales, e.g., non-invasive cancer diagnostics, low-cost plant growth monitoring, and obstacle recognition for autonomous driving (Supplementary Note 8).

In conclusion, we have developed NeuDOT, a novel reconstruction framework that volumetrically resolves the optical absorbance in a scattering medium based on NF. Well-calibrated phantom experiments demonstrated that the proposed method can achieve submillimeter lateral resolution in a single layer and accurately recover complex 3D object to the depth of 14 mm, outperforming the classic DOT method and other DL-based methods. Our work proves that NF can provide an alternative for high-resolution imaging techniques where sophisticated diffuse light propagation dominates the ballistic propagation of photons, paving the ways to more wide and scalable applications in biomedical research, crop monitoring, and autonomous driving.

## Methods

### Basic principle of diffuse optical tomography

Classic DOT reconstruction consists of two steps, i.e., forward modeling and inversion. First, diffuse light propagation and its resulted measurements are calculated by using a physics-based model. Herein we introduce diffusion approximation of Boltzmann transport equation, or diffusion equation (DE) in brief[12]. In CW-DOT, DE can be expressed as:

$$-\nabla \cdot \kappa(r)\nabla\Phi(r) + \mu_a(r)\Phi(r) = 0, \ r \in \Omega \qquad (3)$$

where $\Phi(r)$ denotes the photon density distribution at spatial position $r$ within the scattering volume $\Omega$. $\kappa(r)$ and $\mu_a(r)$ represent the diffusion coefficient and absorption coefficient respectively. A boundary condition is introduced as only photon densities at the surface of the volume can be measured as:

$$\Phi(m) + 2\zeta(c)\kappa(m)\frac{\partial\Phi(m)}{\partial v} = q(m), \ \ m \in \partial\Omega \qquad (4)$$

where $q(m)$ is the source distribution on the domain boundary $\partial\Omega$, $v$ an outward normal, and $\zeta(c)$ a constant accounting for the refractive index mismatch with $c$ denoting the speed of light in the medium. The measurable quantity is the normal photon current across the boundary,

$$\Gamma(m) = -c\kappa(m)\frac{\partial\Phi(m)}{\partial v}, \ m \in \partial\Omega \ . \qquad (5)$$

The unknown parameter, i.e., optical coefficient $\mu_a$, is recovered from the boundary measurement $\Gamma$ by minimizing the objective function Eq. (1).

### Experimental setup

Our CW-DOT imaging system (Fig. 1a) is equipped with a 16-bit CMOS camera (pco.edge 4.2, PCO AG, Kelheim, Germany) and an objective lens (f = 50 mm, Nikon, Tokyo, Japan), providing raw images with 2048 × 2048 pixels and a field-of-view (FOV) of 60 × 60 mm$^2$. A 660 nm wavelength solid-state laser (OBIS, Coherent, CA, US) is used as the CW illumination source featuring a maximum power of 100 mW. The generated free-space laser beam firstly enters the fiber through a coupler (PAF-X-5-B, Thorlabs, NJ, US) and is then directed onto a galvanomirror (basiCube 10, SCANLAB GmbH, München, Germany) with a collimator (F280FC-B, Thorlabs, NJ, US). A transmission type of illumination is implemented, i.e., the outcoming light from galvanomirror reaches the bottom of the phantom whereas the transmitted and scattered light is measured on the top surface. An OD filter (NE01A-A, Thorlabs, NJ, US) and a focus lens (LA1172, Thorlabs, NJ, US, focal length 400 mm) are used to improve the illumination quality with reduced power fluctuation and a light spot with a diameter of less than 1 mm. A 660 nm band-pass filter (FB660-10, Thorlabs, NJ, US) is placed before the camera to reduce the interference of ambient light. We perform a raster scanning strategy with 20 × 20 points on the phantom and capture one image at each scan with the laser power set to 20 mW. The exposure time of each measurement is 500 ms.

MLP construction

To approximate each nodal value of $\mu_a(r)$ in a given mesh with an MLP network, we construct a fully-connected network with eight 256-channel layers, one 128-channel layer and ReLU activations for all 256-channel layers[21] (Fig.5a). Notablly, we perform a skip connection between the input and the fourth layer to faclilate the MLP learning high frequency details and make training more stable. In essence, the MLP network maps each input coordinate $r(x, y, z)$ to its corresponding $\mu_a^\Theta(r)$. In DOT, we focus mostly on regions containing large optical coefficient where high-frequency targets are located, e.g., a high light-absorbing tumor embedded in healthy tissues. To ensure that the MLP best represents high-frequency areas, we apply the positional encoding technique and map each coordinate component $p$ from $\mathbb{R}^1$ space into a high dimensional Fourier-like domain $\mathbb{R}^{2n}$ with encoding function $\gamma(\cdot)$:

$$\gamma(p) = (\sin(2^0 \pi p), \cos(2^0 \pi p), \ldots, \sin(2^{n-1} \pi p), \cos(2^{n-1} \pi p)), p \in \{x, y, z\} \qquad (6)$$

where $n$ denotes the number of different frequencies of the basis function. The original three-dimensional spatial coordinates $r \in \mathbb{R}^3$ are then expanded to a higher dimensional vector $r_e = (\gamma(x), \gamma(y), \gamma(z)) \in \mathbb{R}^{3 \cdot 2n}$, which can be potentially used for learning high-frequency information in the volumetric reconstruction of $\mu_a(r)$. The combination of the MLP and positional encoding enables the network to fit nonlinear scattering light intensity and consequently can model complex distribution of the optical absorption coefficient with a high fidelity.

Afterwards, we minimize the loss function $\Psi(\mu_a^\Theta(r))$ given by Eq. (2) to penalize the dissimilarity between the real measurement $\Gamma$ and the predicted measurement $f(\mu_a^\Theta(r))$ calculated by the combination of MLP and FEM forward model $f$ (Supplementary Note 1). Under this unsupervised setting, we can simply train the MLP via standard gradient back propagation, without pre-trained datasets. Notice that this framework is quite different from previously reported supervised learning methods such as CNN[20]. A few unsupervised learning algorithms such as DeCAF[24] follow a similar framework, but they require additional data to first build a denoising model. Finally, NeuDOT features high flexibility for incorporating various regularization methods to suppress different types of noise in raw measurements.

## Adaptive meshing

Recall that the resolution of DOT reconstruction relies substantially on the accuracy of the forward model. To achieve high resolution, one typically needs to solve Eqs. (3)-(5) on a very fine mesh. In reality, it is computationally infeasible as classical schemes would require very high memory due to the exponential growth of the number of degrees of freedom (DoFs). Similarly, accurately training the NF representation of $\mu_a^\Theta(r)$ in Eq. (2) requires densely evaluating the neural radiance field network at all mesh points, prohibitively expensive even with hierarchical sampling. To balance between computational overhead and numerical accuracy, we further design an adaptive meshing technique and integrate it into the physics-based model and the MLP network (Fig. 5c). This allows us to not only efficiently train NeuDOT on a desired resolution with affordable DoFs but also provides an efficient means to convert between neural and geometric representations. Specifically, we start with a coarse mesh and refine it adaptively and hierarchically by successive loops. During each loop, we employ Gibbs sampling[47] by incorporating the reconstruction solution obtained in the previous loop as a prior information, and use the resulted new points to update the nested mesh (Supplementary Note 3). In particular, we model $\mu_a^\Theta(r)$ in terms of a three-dimensional probability distribution and use the result to conduct volumetric sampling (Fig. 5b). The combination of the positional encoding and adaptive meshing enable NeuDOT to reliably approximate the high-frequency absorption map at high accuracy and resolution, which largely surpasses classical DOT reconstructions and other DL-based methods such asa CNN.

## Phantom experiments

To evaluate the proposed NeuDOT method, we have conducted comprehensive experiments with biomimic phantoms with similar optical properties as realistic biological tissues.

Polydimethylsiloxane (PDMS) (SYLGARD 184, DOW, CA, US) was used as the base of the phantom. We mixed it with titanium dioxide (TiO$_2$, Collins, CA, US) and carbon black powder (CB, Collins, CA, US) to adjust the values of scattering and absorption coefficients respectively. All high-absorption imaging targets were 3D printed using black material (VeroUltra, Stratasys, MN, US). Three phantom experiments were designed for testing different aspects of the proposed method:

**Experiment 1**: The experiment focuses on evaluating the reconstruction performance along the lateral direction by using a optically homogeneous slab phantom with a dimension of $22 \times 22 \times 6.5$ mm$^3$. Different planar patterns featuring high absorbance were placed on the bottom of the phantom as the imaging targets (Fig. 2a). Three patterns were designed: a) a letter of 'T', b) a simplified logo of ShanghaiTech University, and c) two parallel aligned stripes. More specifically, the letter in phantom a) is 14.4 mm in height and 7.1 mm in width with a stroke width of 2 mm. In phantom b), the university logo is composed of two concentric rings (Outer diameter: 18 mm; inner diameter: 13.7 mm; stroke width: 1.2 mm) and a tower in the center (height: 8 mm; width: 4.9 mm). Each stripe in phantom c) is 16.1 mm in height and 1.2 mm in width with a gap of 0.5 mm between these two stripes

**Experiment 2**: We further evaluate reconstruction performance along the axial direction by using a slab phantom with a dimension of $55 \times 55 \times 14$ mm$^3$. The phantom contains three high-absorption rods with different heights. The cross section of each rod is a $6 \times 6$ mm$^2$ square, and heights of these rods are 4 mm, 8 mm, and 12 mm respectively. The base of each rod and the phantom are aligned paralelly (Fig. 3a).

**Experiment 3**: To test a more challenging task that resembles real biological samples, we constructed a 55 × 55 × 14 mm³ cuboid phantom that contains virtual blood vessels. The virtual vessel uses a digital model from the human head cardiovascular system[48] and was rescaled to a bounding box with a dimension of 31 × 22 × 10 mm³. The diameter of the vascular branches ranges from 1.2 mm to 2.5 mm, with the branches located at a depth between z = 3 mm and 13 mm (Fig. 4a)

The optical properties of all the phantoms were carefully calibrated by using a custom-built integrating sphere (2P3/M, Thorlabs, New Jersey, US) with the inverse adding-doubling method[49]. For phantom experiment 1, the estimated background optical properties are $\mu_a$ = 0.108 mm⁻¹ and $\mu_s'$ = 0.529 mm⁻¹; For phantom experiment 2) and 3), the estimated background optical properties are $\mu_a$ = 0.075 mm⁻¹ and $\mu_s'$ = 0.444 mm⁻¹ (Supplementary Note 5).

We used several metrics to quantitatively compare the reconstruction outcomes, including mean square error (MSE), Dice similarity coefficient (DSC), structural similarity (SSIM)), and peak signal-to-noise ratio (PSNR) (Supplementary Note 7). For all experiments, we implemented LM in MATLAB (R2021b, MathWorks, MA, US) with a desktop computer (Intel Xeon Gold 6248R, 128 GB memory) and NeuDOT and CNN using PyTorch[50] on a graphic processor (GeForce RTX 3090, 24 GB memory, NVIDIA, CA, US).

## Acknowledgements

The work is supported by the star-up funding of ShanghaiTech University, National Natural Science Foundation of China (61977047, 61976138, 62105205), and Science and Technology Commission of Shanghai Municipality (21010502400).

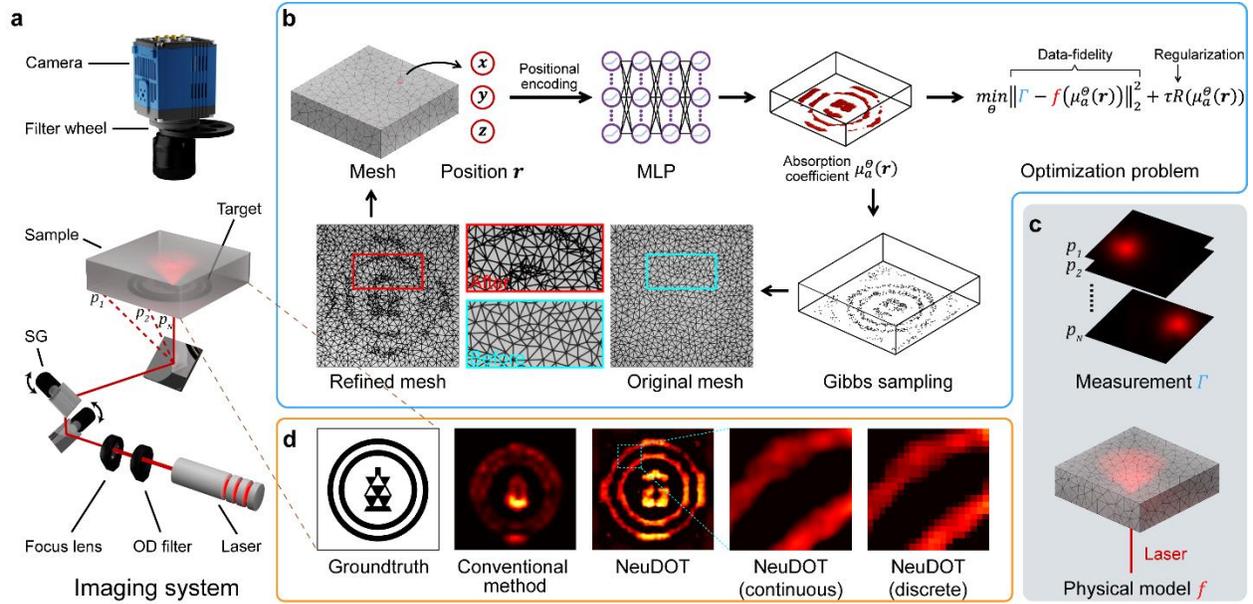

**Fig. 1. Schematic of 3D imaging through scatteing media using NeuDOT. a,** The experimental DOT setup mainly consists of a continuous-wave (CW) laser, a scanning galvanometer (SG) and a CMOS camera. The free-space laser beam raster-scans the bottom of the scattering sample, and the camera captures one image at each illumination point $p_i$ $(i = 1,2,...N)$ as illustrated in (**c**). **b,** A brief pipeline of NeuDOT. Unlike conventional methods where the absorption coefficient $\mu_a$ is stored in a pre-determined grid or mesh, NeuDOT firstly encodes $\mu_a$ by introducing an MLP network ($\mu_a^\Theta(r)$), where its value at an arbitrary position $r$ can be continuously represented. $\Theta$, the parameter of MLP, is subsequently optimized by minimizing an objective function $\Psi(\mu_a^\Theta(r))$ expressed as (2). An FEM-based physical model $f$ is integrated into the NeuDOT pipeline for simulating diffuse light propagation (**c**). We also introduce adaptive meshing by refining the mesh density in the regions of interests during iterations of reconstruction (comparing the content in the red and blue boxes). **d,** The NF-based implicit representation empowers NeuDOT to achieve higher-precision reconstruction through scattering media, compared with conventional methods. Rendering at any desired resolution constitutes another advantage of NeuDOT.

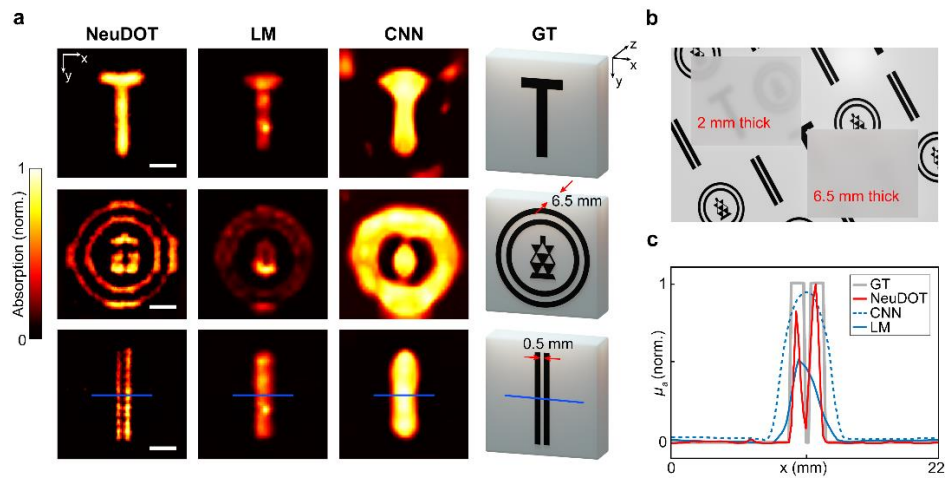

**Fig. 2. Reconstruction performance along the lateral direction**. **a,** Reconstruction results of three different 2D objects generated by NeuDOT, LM, and CNN (scale bar: 4 mm). **b,** Detection of the 2D sample patterns through a thick scattering medium is heavily impeded, as illustrated by phantoms with different thickness. **c,** Line profiles extracted from the central horizontal line across the two-stripe object (**a**, Row 3). NeuDOT achieves high lateral resolution upto 0.5 mm in this case.

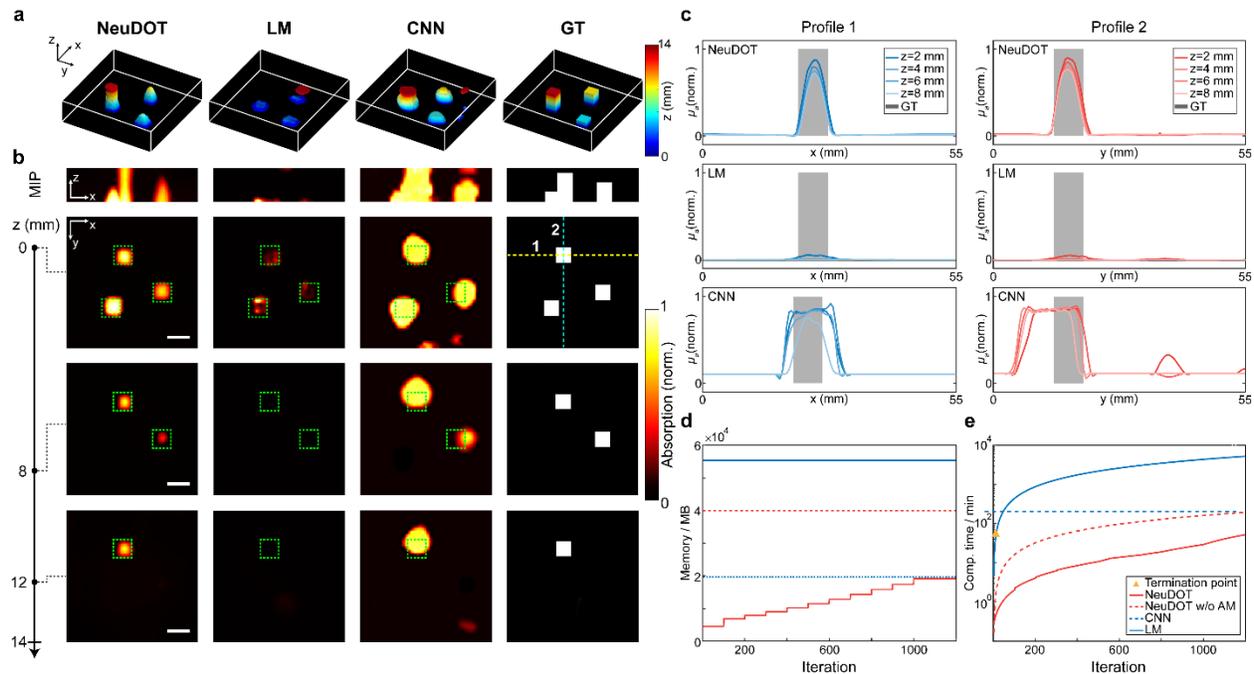

**Fig. 3. Reconstruction performance along the axial direction**. **a,** Volumetric rendering of reconstruction results generated from NeuDOT, LM, and CNN. The last image shows the ground truth (GT) of a slab phantom containing three high-absorbing inclusions with different height (12 mm, 8 mm, 4 mm). **b,** Maximum intensity projection (MIP) and cross sections (at z = 0, 8, 12 mm) for the reconstructed $\mu_a$ maps and the ground truth (scale bar: 10 mm). The real locations of inclusions are indicated with green dotted boxes. **c,** Profile drawings along two lines indicated in **b** show that NeuDOT can recover the position and geometry of each inclusion with higher accuracy than LM and CNN. **d,** Comparison of memory usage. During the iteration, the memory cost of NeuDOT (red solid line) increases incrementally thanks to the step of adaptive meshing. A similar NF-based method using a fixed-dimension fine mesh (red dashed line) to generate the same reconstruction quality as the original NeuDOT will result in 2 times higher memory cost. Both LM (blue solid line) and CNN (blue dashed line) cost higher memory usage compared with NeuDOT. **e,** Comparison of computational time. NeuDOT without adaptive meshing and CNN are both time-consuming (> 200 min). LM converges effectively at the 10th iteration in practice (indicated with a yellow triangle). NeuDOT with adaptive meshing takes comparable computational time as LM (~60 min).

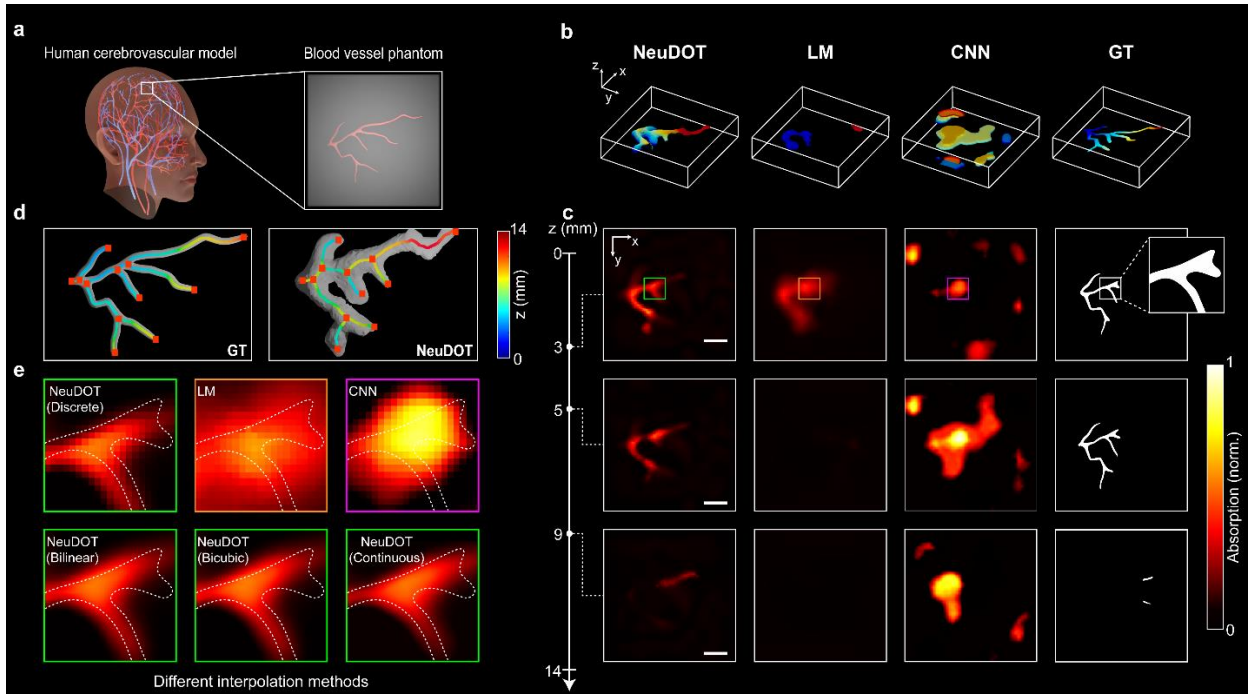

**Fig. 4. Reconstruction results of an artificial vessel model through a scattering medium**. **a**, A virtual vessel (bounding box size: 33×19×9 mm³) is 3D-printed based a human cerebrovascular model[48], and then is embedded in a scattering phantom. **b**, Volumetric reconstruction generated from NeuDOT, LM, and CNN. The last image shows the ground truth (GT) of the vessel-like object. **c**, Cross sections of reconstructed $\mu_a$ maps and GT at the depth z = 3, 5, 9 mm (scale bar: 10 mm). **d**, Comparison of extracted skeletons and endpoints resulted from GT and NeuDOT. NeuDOT is capable of recovering complex 3D structures with well-preserved depth information. **e**, Zoomed views of $\mu_a$ maps resulted from different methods and GT are outlined using different colors (Green: NeuDOT, LM: yellow, CNN: purple). The second row demonstrates that NeuDOT can continuously upsample $\mu_a$ leading to an unprecedented high-resolution result, which is consistent with that of the classic interpolation methods.

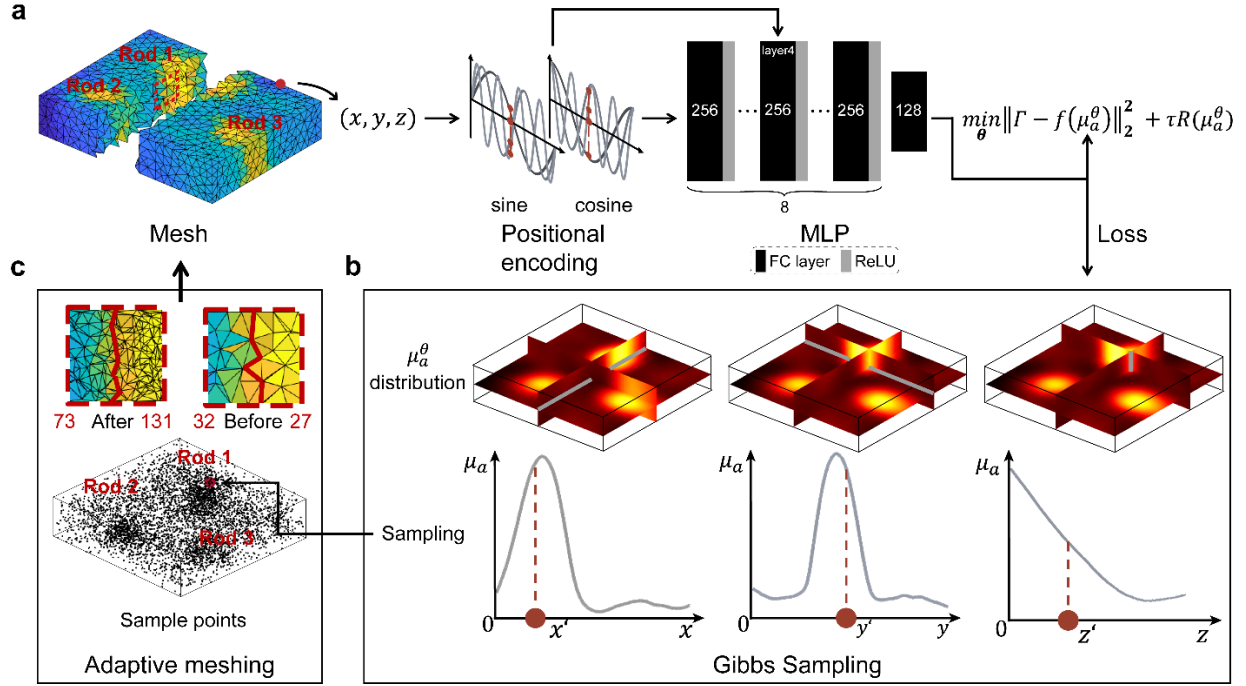

**Fig. 5. A detailed pipeline of NeuDOT with adaptive meshing and Gibbs resampling techniques. a,** NeuDOT aims to reconstruct optical absorption $\mu_a$ by learning an implicit function through an MLP network. Instead of using discrete voxel-based representation, the MLP maps an arbitrary coordinate $r = (x, y, z)$ to its corresponding optical absorption $\mu_a(r)$ in a continuous manner. The imaging object is firstly discretized into a mesh, where the origianal coordinates are encoded via an expanded high-dimensional space. The encoded coordinates are then passed to an MLP where a skip connection between the encoded input and the fourth layer is performed, which is trained by minimizing a loss function. **b,** The output $\mu_a^\theta$ is used as a priori knowledge for resampling by using Gibbs sampling method along three axes. Afterwards, the resampled points are added to the previous mesh for refinement. **c,** The sampled points tend to gather more densely in high-absorbing regions, whilst the refined mesh resolves more structural details at those locations. A phantom containing three rods (Fig. 3) is used for illustration purpose. The reconstructed $\mu_a$ is colour-encoded. The cross section reveals that the edges of the rod becomes more straight after adaptive meshing. The number of nodes on the right side of the rod grows faster than the left side (right: 27 to 131; left: 32 to 73).

Supplementary Information for

# High-resolution tomographic reconstruction of optical absorbance through scattering media using neural fields

Wuwei Ren, Siyuan Shen, Linlin Li, Shengyu Gao, Yuehan Wang, Liangtao Gu, Shiying Li, Xingjun Zhu, Jiahua Jiang, Jingyi Yu

**This file includes:**

# Supplementary Note 1: DOT theory and conventional reconstruction method

The conventional paradigm of DOT reconstruction consists of two parts: 1) a forward FEM solver for simulating the propagation of light in a scattering medium, and 2) a inverse solver that uses an iterative model-based optimization approach to reconstruct the unknown distribution of optical absorption in the volume of interest from boundary measurements of light transmission. Table S1 introduces the notations used in our approach.

## 1.1 Forward modeling and FEM solver

Diffusion equation (DE) is a widely used model to describe the propagation of near-infrared light through a scattering medium. Given a continuous wave (CW) source and a scattering domain $\mathbf{\Omega} \subset \mathbb{R}^3$, bounded by $\partial\mathbf{\Omega}$, the DE can be written as

$$-\nabla \cdot \kappa(\boldsymbol{r})\nabla\Phi(\boldsymbol{r}) + \mu_a(\boldsymbol{r})\Phi(\boldsymbol{r}) = 0, \boldsymbol{r} \in \mathbf{\Omega} \qquad (1)$$

with Robin-type boundary condition

$$\Phi(\boldsymbol{m}) + 2\zeta(c)\kappa(\boldsymbol{m})\frac{\partial \Phi(\boldsymbol{m})}{\partial \nu} = q(\boldsymbol{m}), \boldsymbol{m} \in \partial\mathbf{\Omega} \qquad (2)$$

where the typical value of $g$ in most biological tissues ranges from $0.7$ to $0.95$[1]. The DE is valid under the condition $\mu_a \ll \mu_s'$. The measurable boundary quantity is defined as

$$\Gamma(\boldsymbol{m}) = -c\kappa(\boldsymbol{m})\frac{\partial \Phi(\boldsymbol{m})}{\partial \nu}, \boldsymbol{m} \in \partial\mathbf{\Omega} \qquad (3)$$

Combining Eqs. (2) and (3), we derive the linear relationship between $\Gamma(\boldsymbol{m})$ and $\Phi(\boldsymbol{m})$,

$$\Gamma(\boldsymbol{m}) = \frac{c}{2\zeta(c)}(\Phi(\boldsymbol{m}) - q(\boldsymbol{m})) \qquad (4)$$

For imaging samples involving optical heterogeneity and irregular boundaries in practice, numerical techniques are applied to solve Eqs. (1) and (2). In this work, we use a Galerkin finite element method (FEM) to solve the forward problem. The domain $\Omega$ can be described using a discretized mesh containing $P$ elements and $N$ vertex nodes. Since we only vary the element size and keep the polynomial degree constant, we use the superscript $h$ to denote the finite element mesh basis expansion form. The solution $\Phi(r)$ is approximated by piecewise polynomials over each element,

$$\Phi^h(r) = \sum_{i=1}^{N} \Phi_i u_i(r) \tag{5}$$

where $\Phi^h(r)$ is the numerical approximation of $\Phi(r)$, $\{\Phi_i\}w_{i=1}^{N}$ are the nodal coefficients, $\{u_i(r)\}_{i=1}^{N}$ are basis functions. The coefficients $\kappa(r)$ and $\mu_a(r)$ are expanded in the same manner and denoted by $\kappa^h(r)$ and $\mu_a^h(r)$. Applying integration by parts and boundary condition Eq. (2) on the weak formulation of Eq. (1), we obtain the linear system

$$[K(\kappa^h) + C(\mu_a^h) + \zeta A]\Phi = Q \tag{6}$$

where $\Phi = [\Phi_1, ..., \Phi_N]^T$, $K, C, A$ are sparse symmetric positive definite matrices and given by[2]

$$K_{ij} = \int_{\Omega} \kappa^h(r) \nabla u_i(r) \cdot \nabla u_j(r) dr$$

$$C_{ij} = \int_{\Omega} \mu_a^h(r) u_i(r) u_j(r) dr$$

$$A_{ij} = \int_{\partial\Omega} u_i(m) u_j(m) dm$$

and $Q$ is defined as

$$Q_i = \int_\Omega u_i(r)q(r)dr$$

Since the forward matrix is large for high resolution cases, we employ conjugate gradient method[3] to solve (6). Notice that Eqs. (4)-(6) allow us to predict the measurements $\Gamma(m)$ by absorption coefficient $\mu_a(r)$.

## 1.2 Inversion and regularization

To reconstruct the distribution of the absorption coefficient $\mu_a(r)$ from boundary measurement $\Gamma$, we implement a Levenberg-Marquardt (LM) method[4] to seek the solution of the Tikhonov regularized inverse problem, which is formulated as below:

$$\min_{\mu_a}\|\Gamma - f(\mu_a)\|_2^2 + \tau\|\mu_a - \mu_a^0\|_2^2 \tag{7}$$

where the forward model $f$ is given by Eqs. (1)-(3), $\tau$ is the regularization parameter and $\mu_a^0$ is the prior estimation of the absorption coefficient. In LM iterative procedure, the damping parameter is initially set as 10 and reduced by a factor of $10^{0.25}\rho$ at each iteration, where $\rho$ is the maximum diagonal elements of $J^T J$ and $J$ the Jacobian matrix constructed by adjoint method[5]. More details of LM algorithm for optical tomography can be found in reference[6].

# Supplementary Note 2: CNN-based DOT reconstruction

Convolutional neural network (CNN)-based approaches have recently gained popularity in image reconstruction and shown superior reconstruction performance in DOT[7]. In this work, we compare NeuDOT with a representative CNN-based reconstruction method that adopts multitask learning techniques[8]. Considering the different hardware configuration in our work, we adjust the original procedure of data preparation, when conducting our comparison experiments. The network structure, data preparation and training process are introduced below.

**Network structure:** The framework of the CNN-based method is a residual attention network (Figure. S1a). The first fully connected layer works as a feature extractor by mapping the measurements onto a 9600-column vector. The output of this layer is reshaped into a $40 \times 40 \times 6$ array and delivered to the ensuing convolutional layer with 16 channels and filter size of $3 \times 3 \times 3$. Subsequently, a residual attention block, with 16 channels and 3 different filter sizes (Fig. S1b), is added. The output layer is a convolutional layer with 1 channel and filter size of $3 \times 3 \times 3$, that produces the final reconstruction volume. ReLU is used as the activation function for the entire network.

**Data preparation:** FEM[9] is used as the simulation tool to generate the projection measurements for both training and testing datasets. $20 \times 20$ sources and detectors were placed on the top and bottom sides of the phantom respectively. In Experiment 1, the size of the phantom with 2D objects is set to $22 \times 22 \times 6.5$ mm$^3$. The training data are generated by using a digital phantom containing three ellipsoidal inclusions, which are randomly shaped and distributed within a homogeneous background (Fig. S2). For each inclusion, the three principal radii are uniformly sampled in the ranges of 0.52mm to 12 mm, 0.3 mm to 6.57 mm, and 0.3 mm to 6.57 mm respectively. To make more generic cases, we do not restrict the positioning of the inclusions and allow overlapping of these inclusions. The absorption coefficients $\mu_a$ of the objects and the background are 1 mm$^{-1}$ and 0.1082 mm$^{-1}$ respectively. The scattering coefficient $\mu_s$ is 0.5287 mm$^{-1}$ in the medium. The training set has 4880 samples. For the testing set, we use 1220 samples. In the Experiment 2 and 3, the size of the phantom is set to $55 \times 55 \times 14$ mm$^3$. We generate the data with four randomly distributed ellipsoidal inclusions, with three principal radii uniformly sampled in the ranges of 0.69 mm to 25 mm, 0.4 mm to 13.69 mm, and 0.4 mm to 13.69 mm respectively. $\mu_a$ of the objects and background are 1 mm$^{-1}$ and 0.0748 mm$^{-1}$ respectively, whereas $\mu_s$ is 0.4441 mm$^{-1}$. 7200 training samples and 1800 testing samples are used. Similar to Experiment 1, we have no restrictions on the positioning of these inclusions.

**Training:** The loss in our CNN-based method is the mean square error (MSE) between ground truth and predicted reconstruction. In our experiments, we train the network with batch size of 16 for 200 epochs. We use Adam optimizer[10] with a learning rate $1 \times 10^{-4}$ for experiment 1 ($1 \times 10^{-5}$ for experiment 2 and 3) and hyperparameters $\beta_1 = 0.9$, $\beta_2 = 0.999$ and $\epsilon = 1 \times 10^{-4}$ over the course of optimization.

# Supplementary Note 3: Gibbs sampling

**Gibbs sampling:** As we mentioned in Section 2.3 in the main text, high-resolution DOT reconstruction requires a high-precision forward model, which requires high memory cost. Besides, querying $\mu_a^\Theta(r)$ densely throughout the neural field network in all nodal points is unpractical. In order to use as few points as possible to obtain an accurate forward model, we developed the adaptive meshing technique.

We use Gibbs sampling technique[11] to extract new nodal points for updating the mesh, where the locations of those newly added points contains information of interests. Gibbs sampling is a well-known Markov chain Monte Carlo (MCMC) algorithm to sample from a known multivariate probability distribution. It is proposed based on a fact that for high-dimensional distribution, obtaining a one-dimensional conditional probability distribution is usually easier than the joint distribution. Since NeuDOT reconstructs the $\mu_a^\Theta(r)$ and samples in a 3D volume (unlike NeRF[12] which samples along a single 1D ray), Gibbs sampling is a suitable technique for sampling from a 3D distribution.

The central idea of our sampling strategy is by transforming the 3D sampling task into a combination of three 1D sampling steps. During the sampling procedure, we firstly normalize the probability distribution of $\mu_a^\Theta(r)$ along $x-, y-, z-$ axes. Afterwards, we initialize with a randomized 3D spatial coordinate with the boundary of the mesh $(x^{(0)}, y^{(0)}, z^{(0)}) \sim (x_{boundary}, y_{boundary}, z_{boundary})$. Then we sample along each axis to obtain the next sampling point. This step is the same as the original NeRF application for view synthesis, where the sampling is taken from a single 1D optical ray.

More specifically, we first fix $y^{(0)}, z^{(0)}$ and sample the next sample point $x^{(1)}$ from the conditional probability distribution $p(x^{(0)}|y^{(0)}, z^{(0)})$. As original NeRF, we generate a series of (in NeuDOT we use 10000) uniformly sampled input points $\{x_i^{(0)}, y^{(0)}, z^{(0)} | x_i^{(0)} = \frac{i}{10000} x^{(0)}, i = 1, \ldots, 10000\}$ and pass them to the MLP. Using the network output of $\mu_a^\Theta(r)$, we calculate a cumulative distribution function (CDF) by summation and normalization. Then we get $x^{(1)}$ by uniformly sample from the CDF with inverse sampling. After that, we fix the $x^{(1)}, z^{(0)}$ to sample the $y^{(1)}$ from $p(y^{(0)}|x^{(1)}, z^{(0)})$. At last, we fix the $x^{(1)}, y^{(1)}$ to sample the $z^{(1)}$ from $p(z^{(0)}|x^{(1)}, y^{(1)})$. We repeat the loop three 1D sampling steps and obtain a series of sample points which are located in regions of high interest (high-absorption objects in our case).

**Training parameters:** For balancing the computational cost and reconstruction performance, we sample 1000 new nodal points and update the mesh with an interval of 100 iterations, considering the fact that the combination of sampling and updating mesh for each time takes 1-2 min. We stop adaptive meshing after 1000 iterations until the mesh achieves sufficient fineness. The complete procedure of NeuDOT reconstruction takes 1200 iterations in total. For training parameters, we set the encoding dimension to 4 for positional encoding. We use a L2 norm of $\mu_a^\Theta(r)$ as the regularization term and set hyperparameter $\tau$ to 0.1(see section 2.3 in the main text). We use the Adam optimization with hyperparameters $\beta_1 = 0.9, \beta_2 = 0.999$, and $\epsilon = 1 \times 10^{-8}$. In our experiments, we use a learning rate that begins at $1 \times 10^{-3}$ and decays exponentially to $1 \times 10^{-4}$ through the optimization.

---

**Algorithm** Sampling $n$ points from $\mu_a^\Theta$ distribution

1: **Initialize:**
   $x^0 \sim U(0, x_{boundary})$
   $y^0 \sim U(0, y_{boundary})$
   $z^0 \sim U(0, z_{boundary})$
2: **for** $t = 1$ to $n$ **do**
3:   $x^t \sim p(x^{t-1}|y^{t-1}, z^{t-1})$
4:   $y^t \sim p(y^{t-1}|x^t, z^{t-1})$
5:   $z^t \sim p(z^{t-1}|x^t, y^t)$
6: **end for**

**Algorithm S1. Gibbs Sampling strategy.**

## Supplementary Note 4: Simulation studies

Before conducting experimental reconstruction tasks, we also conduct simulation experiments on a digital phantom. For 2D object cases where the targets are positioned at the bottom of the phantom, the digital phantom is designed as a slab with a dimension of $10 \times 10 \times 6$ mm$^3$ and discretized to 6921 nodes and 23047 elements. For 3D object case where the targets are embedded in the different layers of the phantom, the digital phantom is designed as a box with a dimension of $20 \times 20 \times 6$ mm$^3$ and discretized to 26899 nodes and 152149 elements. We design two 2D object cases containing letters 'ST' (Fig. S3a) and two stripes separated by 0.5 mm (Fig. S3b) and one 3D object case containing two letters placed in different layers with thickness of 1 mm (Fig. S3c). The optical properties of the homogeneous background, $\mu_{a,b}$ and $\mu_{s,b}'$ are set to 0.001 mm$^{-1}$ and 0.99 mm$^{-1}$, respectively. All the embedded inclusions feature high absorption ($\mu_{a,i} = 0.1$ mm$^{-1}$) and same scattering coefficient as the background. We perform the algorithms mentioned in main article to validate the performance of our method (Fig. S3). Only NeuDOT successfully recovers the letters 'ST' and two separated stripes, whereas other methods result in blurry boundaries and distorted shapes of targets. In the 3D object case, NeuDOT clearly reconstructs the letters located at separate depths, whereas other methods may generate reconstruction artefacts (Fig. S3c).

# Supplementary Note 5: Phantom fabrication and calibration

## 5.1 Inverse adding-doubling method

The inverse adding-doubling (IAD) method was used to determine the optical properties of our phantoms. IAD contains two parts: an efficient numerical solver for the radiative transport equation (RTE) (also called adding-doubling method, or AD method), and an optimizer for finding the optimal optical properties to match up the predicted and real values of reflection and transmission. The AD method was introduced for solving RTE specifically in a slab geometry[14]. The quantities that characterize light propagation in a scattering slab are single-scattering phase function $p(\theta)$, albedo $\alpha$, and the optical thickness $\tau$, given by

$$\alpha = \frac{\mu_s}{\mu_s + \mu_a}, \tau = d(\mu_s + \mu_a) \tag{8}$$

with $d$ being the thickness of the slab. The phase function can be approximated by a Henyey-Greenstein function[15] given by:

$$p(v) = \frac{1}{4\pi} \frac{1 - g^2}{(1 + g^2 - 2gv)^{3/2}} \tag{9}$$

where $v$ is the cosine of the scattering angle ($v = \cos\theta$). The quantities of interest are reflection and transmission, which are both normalized to the irradiance on the sample surface and ranges between 0 and 1, with 1 representing full transmission. The total reflection $R_T$ is the sum of light that the sample specularly reflected and backscattered, and the total transmission $T_T$ is the sum of light that passes through the sample (including the light traveling through the sample without being scattered). For a non-absorbing sample, $R_T + T_T = 1$. AD method calculates the reflection and transmission values under the following assumptions: 1) the light source is steady-state and unpolarized; 2) The sample is homogeneous; 3) the sample geometry is considered as an infinite plane-parallel slab of finite thickness; 4) the sample has a uniform refractive index. Once the reflection $R(v, v')$ and transmission $T(v, v')$ for light incident at angle $v$ and exiting at angle $v'$ for one slab are known, the reflection and transmission of a slab that is twice as thick can be found by stacking two identical slabs and adding the two values of each slabs[16]. Then we can calculate the reflection and transmission for an arbitrary slab first by finding the reflection and transmission for a thin slab with the same optical properties and then by doubling until the thickness becomes the same. The inverse problem is to estimate the albedo, the optical path, and the anisotropy given a set of reflection and transmission measurements. For a particular set of optical properties, the distance between the predicted values and real measurements are defined as a sum of relative errors:

$$M(\alpha, \tau, g) = \frac{|R_{pred} - R_{meas}|}{R_{meas} + 10^{-6}} + \frac{|T_{pred} - T_{meas}|}{T_{meas} + 10^{-6}} \tag{10}$$

The problem can be solved using an iteration method based on downhill simplex method of Nelder and Mead[17]. The value of $g$ is usually known from the literature and once we solve $\alpha, \tau$, we can obtain $\mu_a, \mu_s$ using Eq. (8).

## 5.2 Calibration of phantoms

We calibrated two phantoms in our experiment: a $55 \times 55 \times 14$ mm³ phantom A and a $22 \times 22 \times 6.5$ mm³ phantom B (Fig. S4a). The calibration system consists of a integrating sphere (2P3/M, Thorlabs, NJ, US), a 660 nm wavelength solid-state laser (OBIS, Coherent, CA, US), and an optical power meter (PM100D, Thorlabs, NJ, US). The measurements are modified as

$$M_R = r_{std} \cdot \frac{R(r_s^{direct}, r_s) - R(0,0)}{R(r_{std}, r_{std}) - R(0,0)}, M_T = \frac{T(t_s^{direct}, r_s) - T_{dark}}{T(0,0) - T_{dark}} \tag{11}$$

where $R(r_s^{direct}, r_s), T(t_s^{direct}, r_s)$ are the reflection and transmission of phantom respectively, $R(0,0), T(0,0)$ are the reflection and transmission of air, $R(r_{std}, r_{std})$ is the reflectance of standard sample, and $T_{dark}$ is the transmission of ambient light in dark[18]. The transmittance and reflectance of phantom measured by integrating sphere are shown in Table.S2 and we also measured other parameters that the IAD algorithm requires such as the refractive index of the sample, the thickness of the sample, entrance port diameter of the integrating sphere (Fig. S4b). The estimated optical parameters are given in Table. S3.

# Supplementary Note 6: Additional phantom experiment with two letters

An additional phantom experiment was performed to further test our proposed method. In this case, a 55 × 55 × 14 mm$^3$ slab phantom embedded with two high-absorption letters ('S' and 'T') was used. The bounding box size for 'S' and 'T' is 25 × 18 × 4 mm$^3$ and 24 × 17 × 4 mm$^3$, respectively. Besides, 'S' is positioned on the bottom, the distances between 'S' and 'T' are 4 mm vertically and 7 mm horizontally (Fig. S5a). NeuDOT is shown to recover the height of two letters more accurately than the other two methods. The shape of two letters are fully recovered by NeuDOT, whereas the other two methods result in blurry boundaries. NeuDOT reconstruction generates slight artifacts in the axial direction of letter 'T', which also appears in the case of three rods. Nevertheless, NeuDOT still achieves superior performance compared with CNN and LM methods.

.
# Supplementary Note 7: Quantitative evaluation of reconstruction algorithms

We compare the computational cost of all methods mentioned in the main article. Additionally, we test NeuDOT without adaptive meshing (Table. S4). The FEM meshes used in our experiments, the memory usage and reconstruction time of each method are listed in Table. S4.

To quantitatively analyze the reconstruction results of each method, we introduce four indicators: mean squared error (MSE), Dice similarity coefficient (DSC), structural similarity (SSIM), and peak signal-to-noise ratio (PSNR). The input is a normalized 3D array $A$ representing the reconstructed volume and a normalized 3D array $B$ representing the ground truth. Each indicator is calculated as follows

$$MSE = \frac{1}{N}\sum_i^N (A_i - B_i)^2 \tag{12}$$

$$SSIM = \frac{(2\mu_A\mu_B + C_1)(2\sigma_{AB} + C_2)}{(\mu_A^2 + \mu_B^2 + C_1)(\sigma_A^2 + \sigma_B^2 + C_2)} \tag{13}$$

$$PSNR = 10\log_{10}\frac{peakval^2}{MSE} \tag{14}$$

$$DSC = 2 * \frac{|intersection(A_b, B_b)|}{|A_b| + |B_b|} \tag{15}$$

where $peakval$ is set to 1 since the volumes are normalized and $\mu_A, \mu_B, \sigma_A, \sigma_B, \sigma_{AB}$ are the mean, standard deviation, and cross-covariance of array $A$ and $B$ respectively. The binarized volumes $A_b$ and $B_b$ are created by thresholding: the 2D volumes are threshold by 10% of the maximum value and the 3D volumes are threshold by 2.5 times of the mean.

# Supplementary Note 8: Future applications of NeuDOT

The NeuDOT has shown its ability to produce high quality reconstruction, e.g., the structure of small and complex blood vessel target. A lot of subsequent work of neural fields shows the possibility to further speed training of NeuDOT (as mentioned in main text). We believe NeuDOT can been applied to broader applications (Fig. S6) such as 1) Non-invasive and high-resolution breast cancer diagnostics, which involves no ionizing radiation. 2) Obstacle recognition in automated driving on foggy days. 3) Plant root growth monitoring, which needs to extract accurate structure of root through turbid fluid.

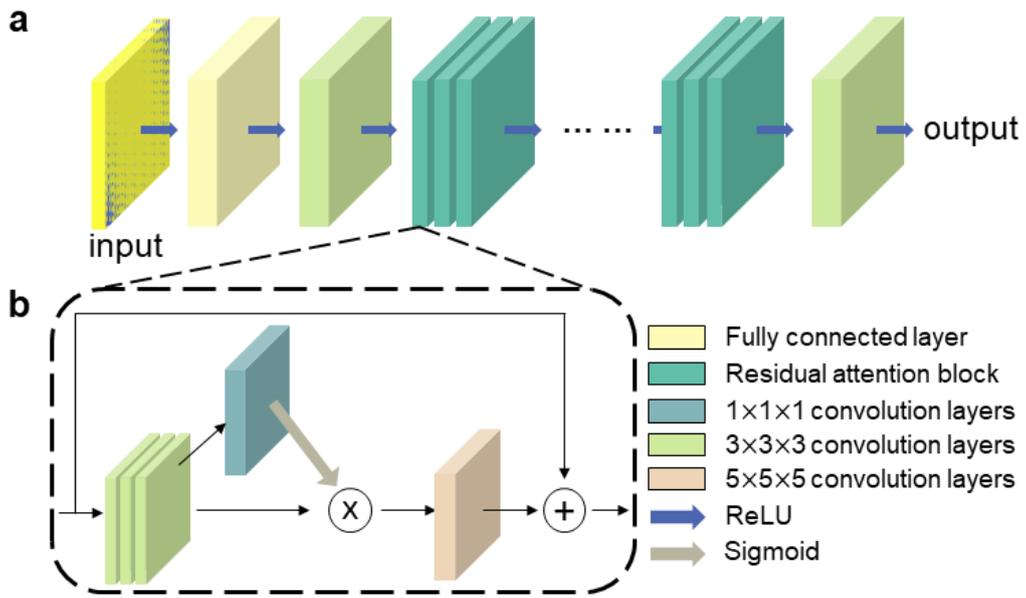

**Fig.S1 The network architecture of the CNN-based reconstruction method. a** Structure of the CNN framework. **b** Structure of the residual attention block.

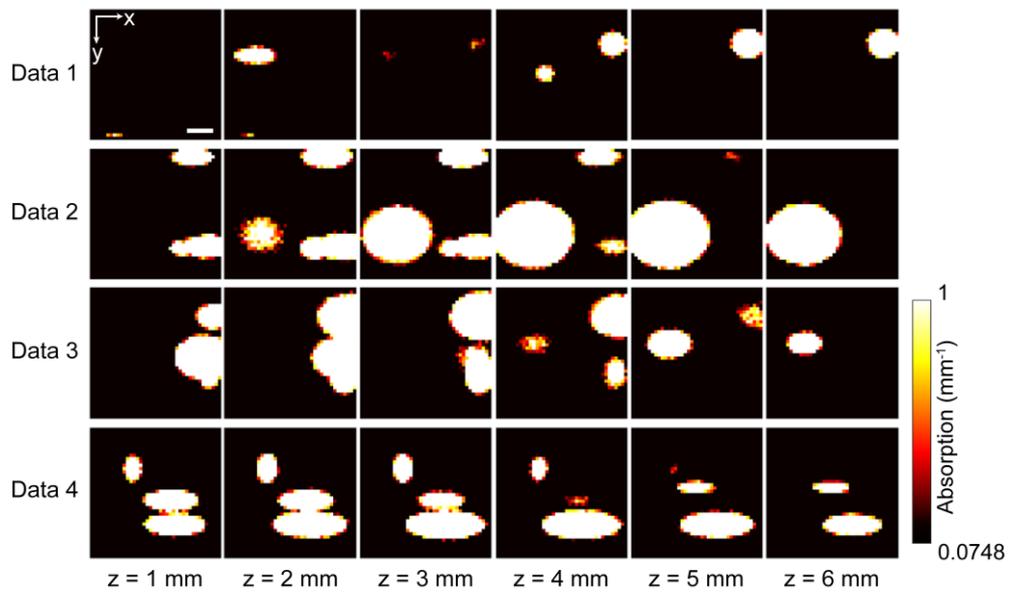

**Fig.S2 Examples of training data for the CNN-based method for Experiment 1.** Ground truth of $\mu_a$ at the different depth. The scale bar indicates a length of 11 mm.

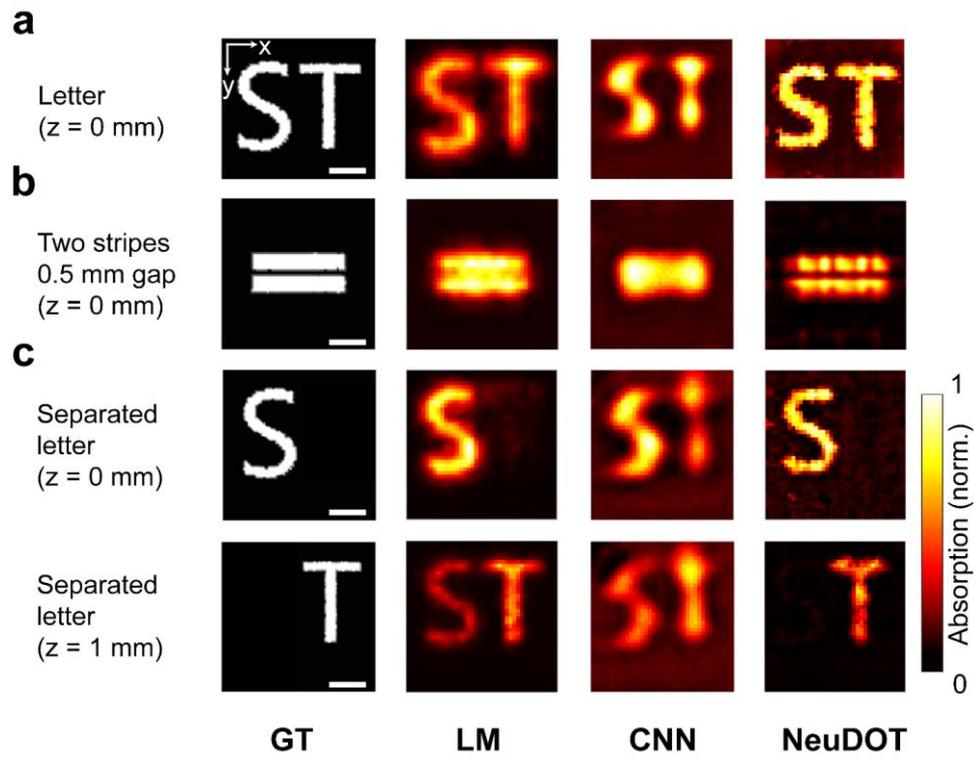

**Fig.S3** Reconstruction results of **simulation studies.** (Scale bar: 5 mm)

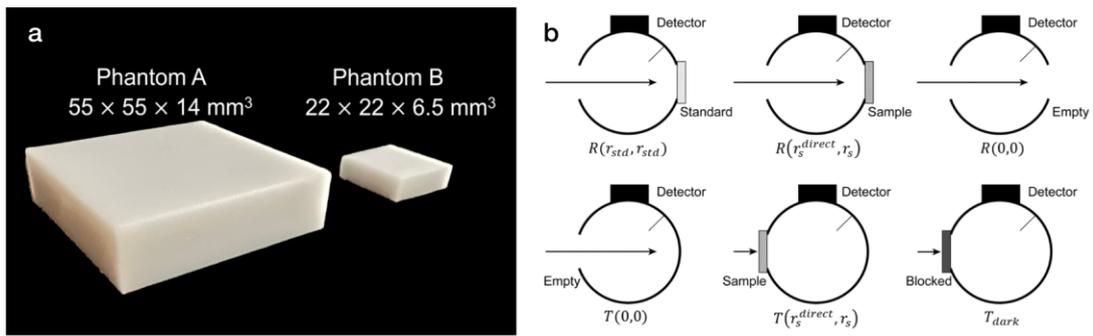

**Fig.S4 Silicone phantoms and calibration procedure. a** Two phantoms for experiments. **b** the incident light (arrow) enters the integrating sphere (arc), and different quantities are recorded by the power detector on the top (black box).

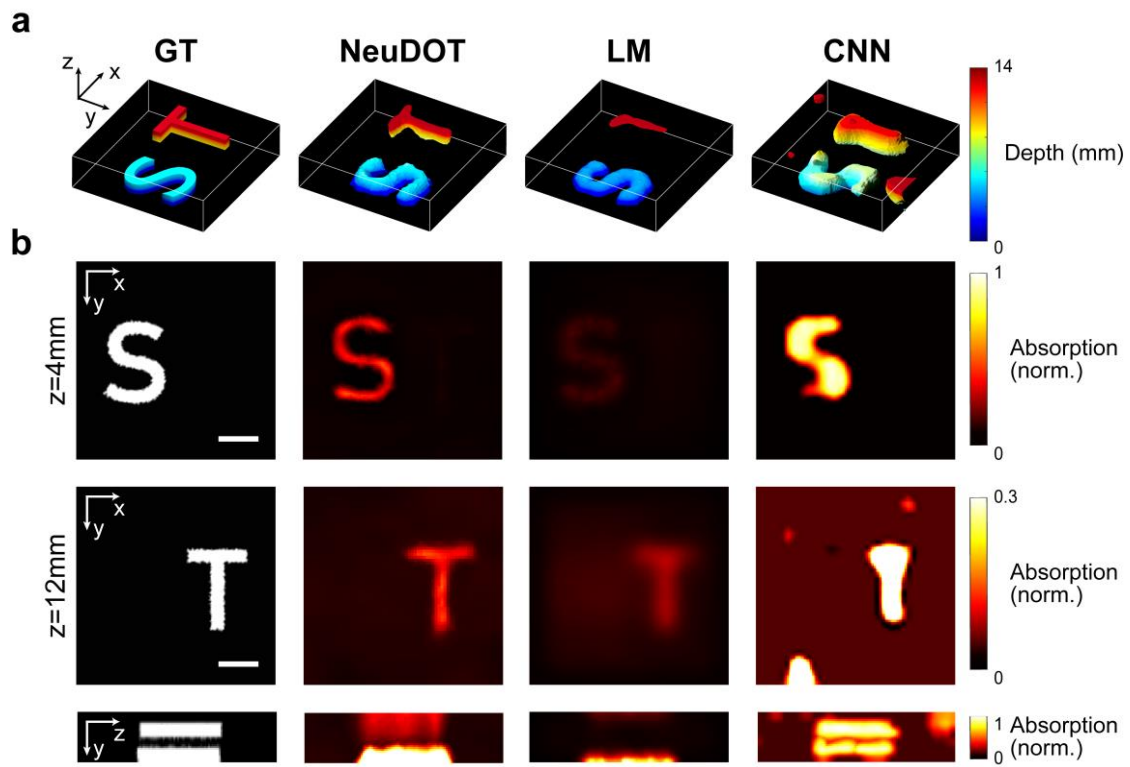

**Fig.S5 Reconstruction results of letters 'ST' embedded in the phantom**. **a** 3D illustration of target and reconstruction results from three methods. The reference thickness of two letters are both 4 mm. 'S' is located above the plane z=0 mm and 'T' is located above the plane z=8 mm. **b** Axial views at z = 4 and z = 12 mm. Scale bar: 10 mm. The maximum intensity projection on yz-plane is shown at the bottom, highlighting better depth accuracy compared to LM and CNN

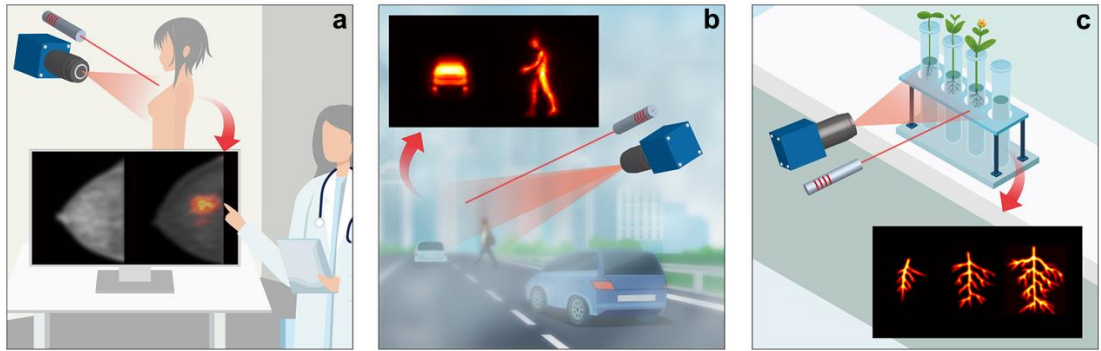

**Fig.S6 Illustration of future applications of NeuDOT. a** Breast cancer diagnostics. **b** Automated driving. **c** Plant monitoring.

| Symbol | Quantity | Unit |
|---|---|---|
| $\Omega$ | Scattering domain $\Omega \subset \mathbb{R}^3$ | mm |
| $\partial\Omega$ | Boundary of $\Omega$ | mm |
| $\boldsymbol{r}$ | Position vector $\boldsymbol{r} = (x, y, z) \in \Omega$ | mm |
| $\Phi(\boldsymbol{r})$ | Photon density distribution at $\boldsymbol{r}$ | #photons mm$^{-3}$ |
| $\mu_a(\boldsymbol{r})$ | Absorption coefficient at $\boldsymbol{r}$ | mm$^{-1}$ |
| $\mu_s(\boldsymbol{r})$ | Scattering coefficient at $\boldsymbol{r}$ | mm$^{-1}$ |
| $g$ | Anisotropy factor | NA |
| $\mu_s'(\boldsymbol{r})$ | Reduced scattering coefficient at $\boldsymbol{r}$, $\mu_s'(\boldsymbol{r}) = (1-g)\mu_s$ | mm$^{-1}$ |
| $\kappa(\boldsymbol{r})$ | Diffusion coefficient at $\boldsymbol{r}$, $\kappa(\boldsymbol{r}) = 1/3(\mu_a + \mu_s')$ | mm |
| $\boldsymbol{m}$ | Measurement position vector constrained to a surface $\partial\Omega$ | mm |
| $q(\boldsymbol{m})$ | Source distribution on boundary $\partial\Omega$ | #photons mm$^{-2}$ |
| $\nu$ | Outward normal at $\partial\Omega$ | NA |
| $c$ | Speed of light in the medium | mm s$^{-1}$ |
| $\zeta(c)$ | Boundary mismatch | NA |
| $\Gamma$ | Measurable quantity | NA |

**Table.S1** Notation used throughout this section.

|  | $R(r_{std}, r_{std})(\mu w)$ | $R(r_s^{direct}, r_s)(\mu w)$ | $R(0,0)(\mu w)$ |
|---|---|---|---|
| **Phantom A** | 18.6 | 2.34 | 0.03 |
| **Phantom B** | 18.2 | 2.23 | 0.06 |
|  | $T(0,0)(\mu w)$ | $T(t_s^{direct}, r_s)(\mu w)$ | $T_{dark}(\mu w)$ |
| **Phantom A** | 18.6 | 0.07 | 0 |
| **Phantom B** | 18.2 | 0.74 | 0 |

**Table.S2 Measured raw data of using integrating sphere.**

|  | $\mu_a$(mm⁻¹) | $\mu_s'$(mm⁻¹) |
|---|---|---|
| **Phantom A** | 0.0748 | 0.4441 |
| **Phantom B** | 0.1082 | 0.5287 |

**Table.S3 Calibrated optical properties.**

| Target | FEM mesh | | Memory cost / MB | | | |
|---|---|---|---|---|---|---|
| | nodes | elements | LM | NeuDOT | NeuDOT w/o AM | CNN |
| Letter 'T' | 21304 | 119018 | 13312 | 18736 | 38753 | 22180 |
| Logo | | | | | | |
| Two stripes | | | | | | |
| Letter 'ST' | 22279 | 124024 | 55421 | 19060 | 41245 | 19490 |
| Rods | | | | | | |
| Blood vessel | | | | | | |
| Target | Reconstruction time / min | | | |
|---|---|---|---|---|
| | LM | NeuDOT | NeuDOT w/o AM | CNN |
| Letter 'T' | 14 | 46 | 203 | 122 |
| Logo | | | | |
| Two stripes | | | | |
| Letter 'ST' | 55 | 60 | 219 | 210 |
| Rods | | | | |
| Blood vessel | | | | |

**Table.S4 Memory cost and reconstruction time of LM, CNN and NeuDOT.**

|        | Method | 2D Experiments | | |
|--------|--------|----------------|---|---|
|        |        | Letter 'T' | Logo | Two stripes |
| MSE  | LM     | 0.0047 | 0.0298 | 0.0057 |
|      | NeuDOT | **0.0014** | **0.0292** | **0.0061** |
|      | CNN    | 0.0177 | 0.0474 | 0.0069 |
| DSC  | LM     | 0.7689 | 0.3818 | 0.4758 |
|      | NeuDOT | **0.8297** | **0.4387** | **0.8067** |
|      | CNN    | 0.3619 | 0.3352 | 0.3393 |
| SSIM | LM     | 0.5836 | 0.2824 | 0.2063 |
|      | NeuDOT | **0.7684** | **0.5639** | **0.2201** |
|      | CNN    | 0.0384 | 0.0858 | 0.1020 |
| PSNR | LM     | 23.3235 | 15.2600 | **22.4458** |
|      | NeuDOT | **27.7593** | **15.3505** | 22.1653 |
|      | CNN    | 17.5094 | 13.2408 | 21.6363 |
|        | Method | 3D Experiments | | |
|        |        | Letter 'ST' | Blood vessel | Rods |
| MSE  | LM     | 0.0198 | 0.0093 | 0.0203 |
|      | NeuDOT | **0.0154** | **0.0063** | **0.0091** |
|      | CNN    | 0.0287 | 0.0249 | 0.0437 |
| DSC  | LM     | 0.4499 | 0.0602 | 0.3974 |
|      | NeuDOT | **0.6117** | **0.2048** | **0.7834** |
|      | CNN    | 0.4906 | 0.1930 | 0.4299 |
| SSIM | LM     | 0.7076 | 0.6553 | 0.6181 |
|      | NeuDOT | **0.8018** | **0.7054** | **0.8736** |
|      | CNN    | 0.7201 | 0.4380 | 0.6223 |
| PSNR | LM     | 17.0230 | 20.2952 | 16.9263 |
|      | NeuDOT | **18.1268** | **22.0394** | **20.3868** |
|      | CNN    | 15.4203 | 16.0346 | 13.5972 |

**Table.S5 Quantitative evaluation.** The metrices indicating highest performance are highlighted in bold.